\documentstyle[sprocl,epsf]{article}

\def\Journal#1#2#3#4{{#1} {\bf #2}, #3 (#4)}

\def\NPB{{\em Nucl. Phys.} B}
\def\NPBPS{{\em Nucl. Phys.} B (Proc. Suppl.)}

\def\PRL{\em Phys. Rev. Lett.}
\def\PRD{{\em Phys. Rev.} D}

\newcommand{\la}[1]{\label{#1}}
\newcommand{\be}{\begin{equation}}
\newcommand{\ee}{\end{equation}}
\newcommand{\ba}{\begin{eqnarray}}
\newcommand{\ea}{\end{eqnarray}}
\newcommand{\bi}{\begin{itemize}}
\newcommand{\ei}{\end{itemize}}

\newcommand{\fig}{Fig.~}

\newcommand{\tr}{{\rm Tr\,}}

\newcommand{\fr}[2]{{\frac{#1}{#2}}}

\newcommand{\bfx}{{\bf x}}

\def\lsi{\raise0.3ex\hbox{$<$\kern-0.75em\raise-1.1ex\hbox{$\sim$}}}
\def\gsi{\raise0.3ex\hbox{$>$\kern-0.75em\raise-1.1ex\hbox{$\sim$}}}

\newcommand{\gsim}{\mathop{\gsi}}

\begin{document}

\title{THE PROPERTIES OF THE CRITICAL POINT 
IN ELECTROWEAK THEORY%
\footnote{Presented at SEWM'98, Copenhagen, 2.-5.12.1998. 
Based on~Ref.~\cite{endpoint}.}}

\author{M. TSYPIN}

\address{Department of Theoretical Physics, Lebedev Physical Institute, \\
117924 Moscow, Russia}

\maketitle\abstracts{
We study the properties of the electroweak theory in the vicinity
of the critical point --- the endpoint of the first order 
phase transition line that exists in a wide class of theories, including
the Standard Model and a part of the parameter space of the
Minimal Supersymmetric Standard Model.
We find that the critical point belongs to the 3d Ising universality class.
The endpoint in the Standard Model is located at the 
Higgs mass below the existing experimental lower bound, 
implying that there is no electroweak phase transition in this theory.
}

The properties of the electroweak theory at temperature of order 
100 GeV, including the existence, the order and other properties
of the phase transition, attract considerable attention, 
as they play an important role in our understanding
of the baryogenesis in the early Universe \cite{RSreview}.

The phase structure of the Standard Model 
(\fig\ref{phdiag}, see \cite{isthere,karsch,gurtler1,fodor})
has the following characteristic features:

%%%%%%%%%%%%%%%%%%%%%%%%%%%%%%%%% FIGURE
\begin{figure}
 
\epsfxsize=7cm
\centerline{\epsffile{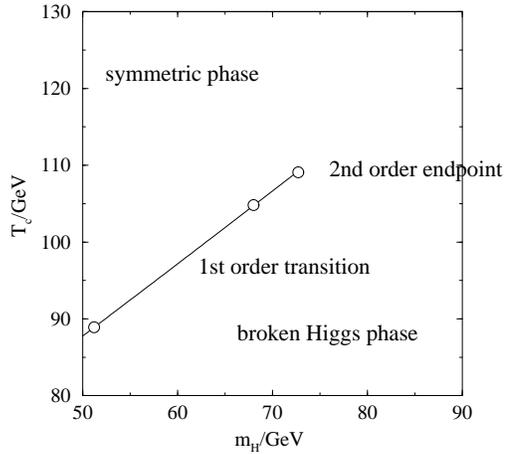}}
 
\caption{The phase diagram of the Minimal Standard Model.\label{phdiag}}

\end{figure} 
%%%%%%%%%%%%%%%%%%%%%%%%%%%%%%%%%%%%

The system has two parameters: the Higgs mass $m_H$ and 
temperature $T$.

There is a first order phase transition line that ends in an endpoint.
The jumps of observables decrease towards the endpoint and finally vanish 
there.

The study of correlation functions show that near the endpoint
one scalar mode becomes light:
\begin{equation}
m_{0^{++}} \ll \mbox{other~masses}.
\end{equation}
The corresponding correlation length diverges: $\xi_{0^{++}} \to \infty$.

Thus one expects the endpoint to be a point of a second order
phase transition (a critical point), and the interesting question
is, to which universality class does it belong. The phase 
diagram in \fig\ref{phdiag} looks quite similar to the 
phase diagram of the usual liquid-gas system in the pressure vs.
temperature plane, in which case it is known that the critical
point belongs to the universality class of the 3d Ising model
(see, e.g., \cite{wilding}). We have performed a detailed 
study \cite{endpoint} of the properties of the electroweak theory 
at the critical point, and our conclusion is that this endpoint
also belongs to the 3d Ising model universality class.

%%%%%%%%%%%%%%%%%%%%%%%%%%%%%%%%% FIGURE
\begin{figure}

\centerline{
\epsfysize=4.8cm \epsfbox[36 40 539 468]{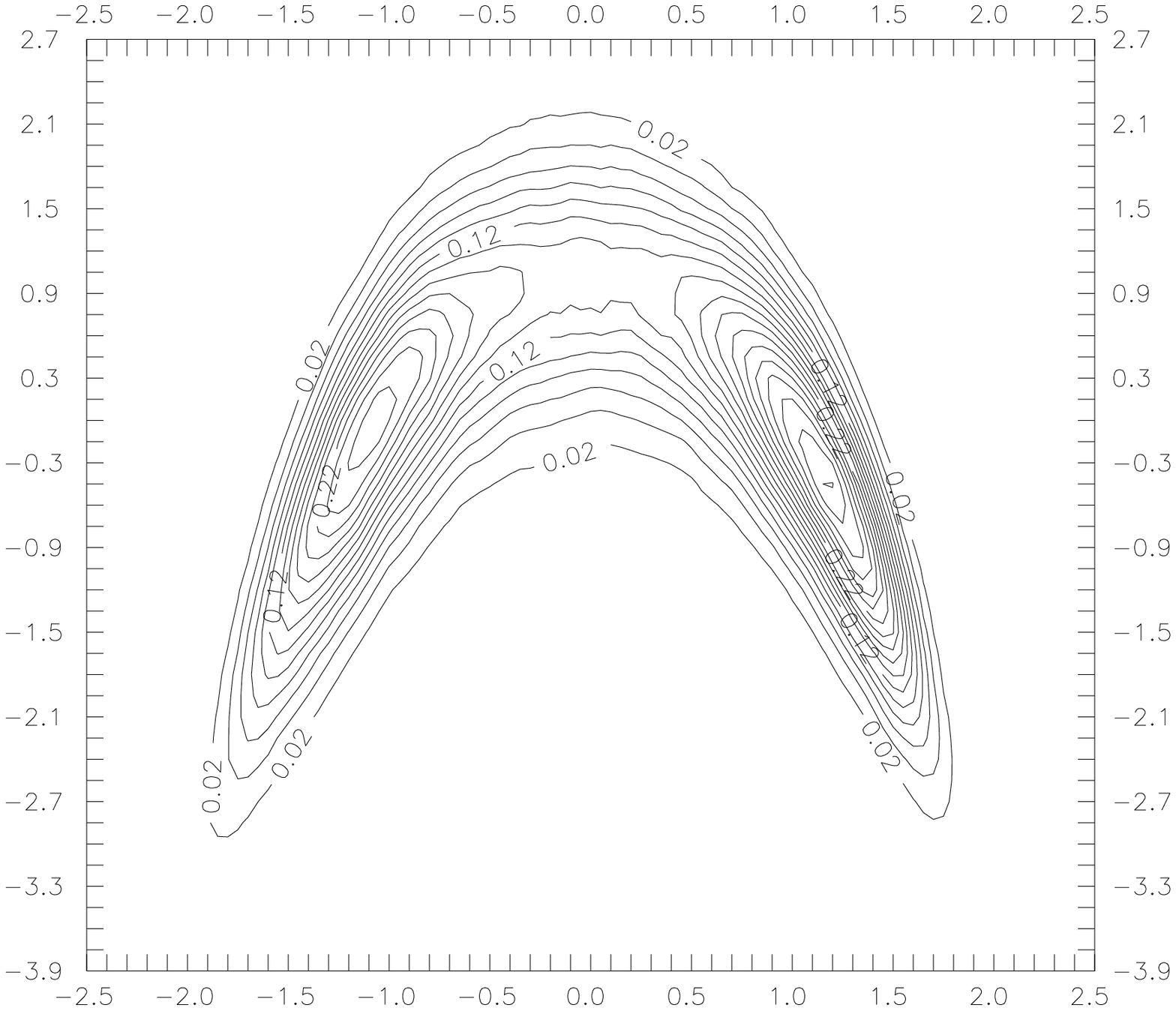}
\hspace*{5mm}
\epsfysize=4.8cm \epsfbox[36 40 539 468]{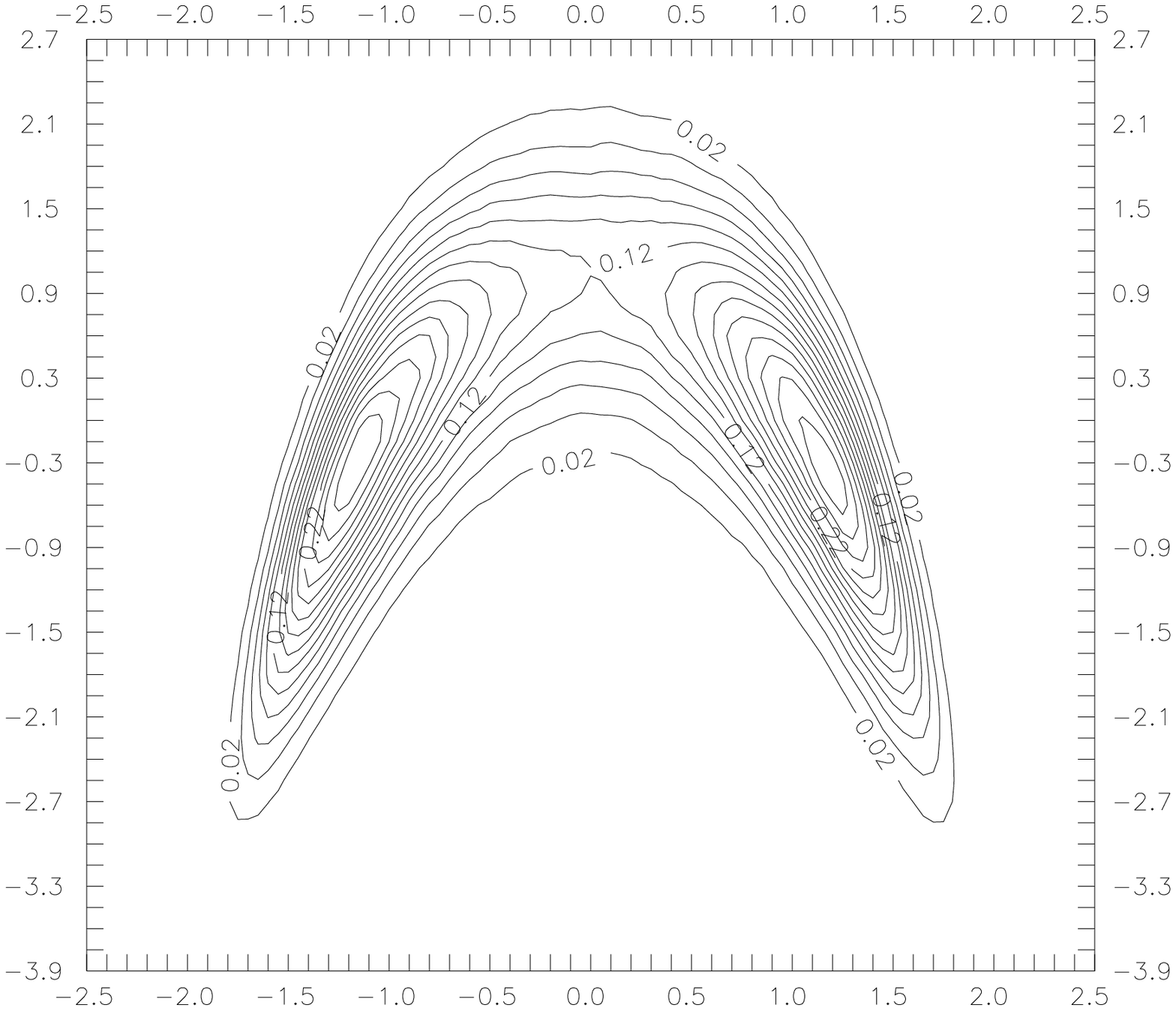}
}
\centerline{(a) \hspace*{5.5cm} (b)}
\vspace*{2mm}

\centerline{
\epsfysize=4.8cm \epsfbox[36 40 539 468]{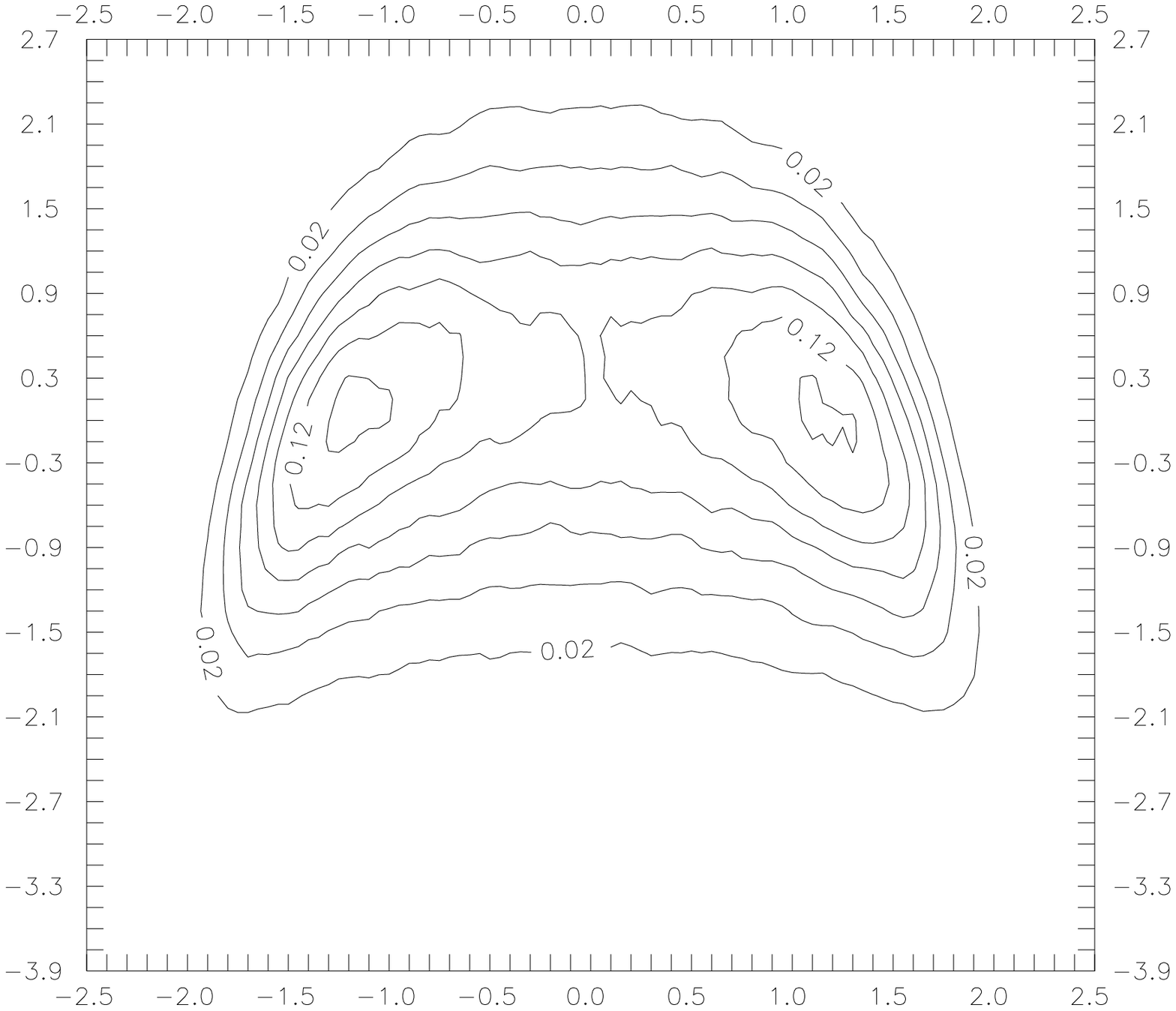}
\hspace*{5mm}
\epsfysize=4.8cm \epsfbox[36 40 539 468]{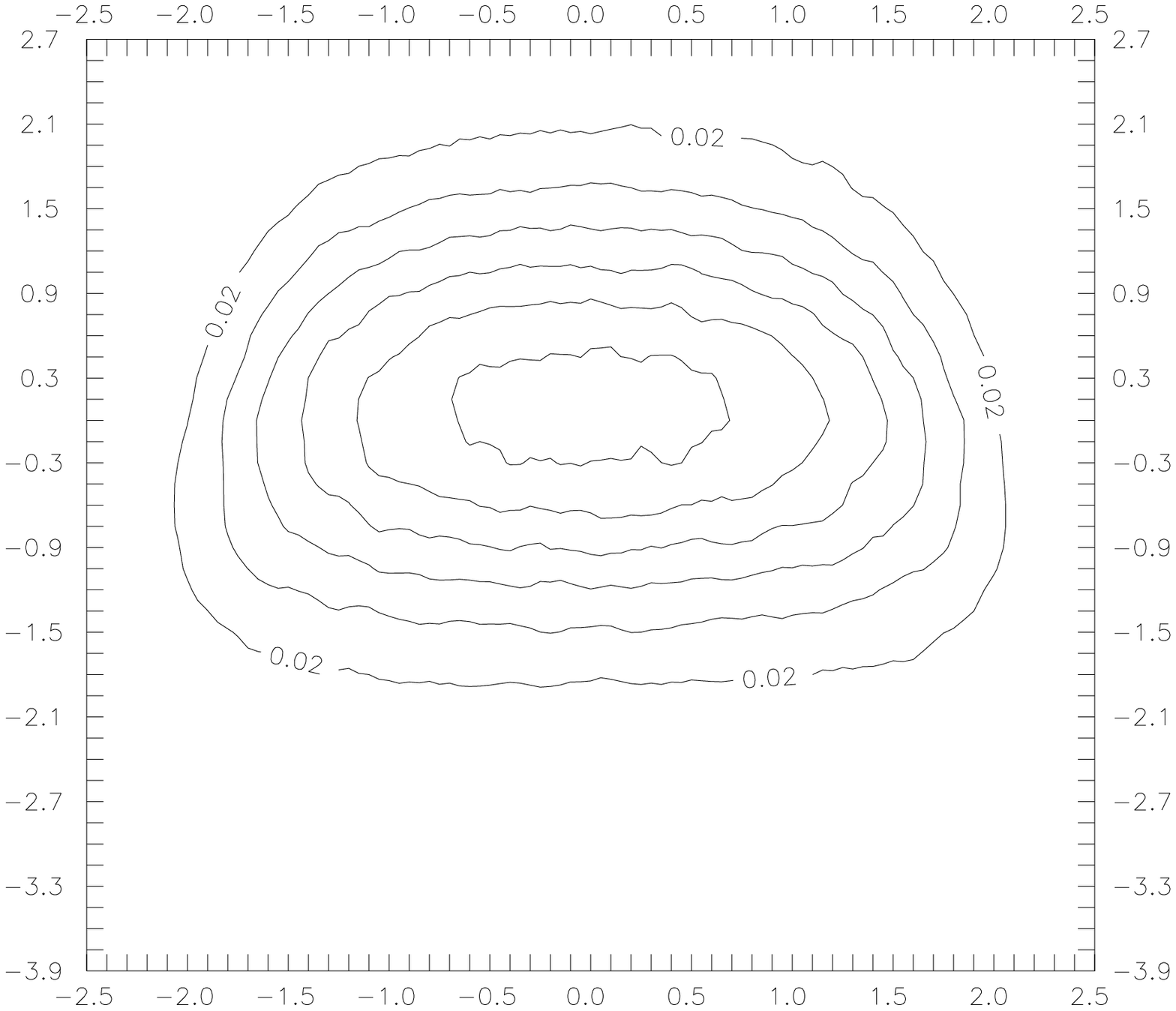}
}
\centerline{(c) \hspace*{5.5cm} (d)}
\vspace*{2mm}

\caption{
The smoothed and normalized probability distributions,
at the critical point, for
(a) the model (\protect\ref{latticeaction}) at the volume $64^3$,
(b) the Ising model at the volume $58^3$,
(c) the O(2) spin model at $64^3$,
(d) the O(4) spin model at $64^3$.
The $x$-axis is the magnetic direction and
the $y$-axis the energy direction. \la{2dplots}}
\end{figure}
%%%%%%%%%%%%%%%%%%%%%%%%%%%%%%%%%%%%

We work within the dimensional reduction approach. After
integrating in the partition function of the full theory
over the Euclidean time-dependent modes we obtain the
three-dimensional effective theory:
\be
S  =  \int\! d^3x \Bigl[
\frac{1}{4}\tr F^a_{ij}F^a_{ij}+
(D_i\phi)^{\dagger}(D_i\phi)+
m_3^2\phi^{\dagger}\phi+
\lambda_3 (\phi^{\dagger}\phi)^2\Bigr],
\label{action}
\ee
which depends on the dimensionful gauge coupling $g_3$ and two 
dimensionless ratios
\be
x=\lambda_3/g_3^2, \quad
y=m_3^2(\mu)/g_3^4, \label{xy}
\ee
where $m_3^2(\mu)$ is the renormalized mass parameter.
This effective theory is put on a lattice. The corresponding
lattice action reads
\begin{eqnarray}
S&=& \beta_G \sum_\bfx \sum_{i<j}(1-\fr12 \tr P_{ij}) 
 - \beta_H \sum_\bfx \sum_i
\fr12\tr\Phi^\dagger(\bfx)U_i(\bfx)\Phi(\bfx+i)
\nonumber \\
 &+& \sum_\bfx
\fr12\tr\Phi^\dagger(\bfx)\Phi(\bfx) + \beta_R\sum_\bfx
 \bigl[ \fr12\tr\Phi^\dagger(\bfx)\Phi(\bfx)-1 \bigr]^2
\la{latticeaction}. 
%\\
%&\equiv& S_G+S_\rmi{hopping}+S_{\phi^2}+S_{(\phi^2-1)^2}.
%\nonumber
\end{eqnarray}
At any fixed $\beta_G$ (we use $\beta_G = 5, 8, 12$; continuum
limit corresponds to $\beta_G \to \infty$) there are two parameters,
$\beta_H$ and $\beta_R$. To characterize the critical point, we have to 
solve the following main problems:

1. Locate the critical point in the plane $(\beta_H, \beta_R)$.

2. Find the observables that show the same scaling behavior
as magnetization and energy in the Ising model ($M$-like and 
$E$-like observables).

3. Find the critical indices.

In practice, a straightforward approach to these problems,
which would imply performing the simulations in the vicinity
of the endpoint and studying the power-law singularities
in dependences of observables on the distance from the endpoint,
runs into severe technical difficulties with the lattice size,
because the whole hierarchy of scales has to fit into the lattice:
\be
1 \ll \xi_{heavy} \ll \xi_{0^{++}} \ll L.
\ee
The first inequality, $1 \ll \xi_{heavy}$, is controlled by 
$\beta_G$ and means that the lattice theory (\ref{latticeaction})
should be sufficiently close to its continuum limit.
The second one, $ \xi_{heavy} \ll \xi_{0^{++}}$, is controlled
by the distance from the critical point: if it is not 
satisfied, large deviations from scaling are unavoidable.
Finally, the inequality $\xi_{0^{++}} \ll L$ is to be 
satisfied to suppress the finite volume effects (in practice,
$L/\xi \gsim 5$ is required for this purpose), and this leads
to a prohibitively large lattice size.

Thus we have adopted the finite size scaling approach, which
makes it possible to {\em use} finite size effects
(that is, extract the necessary information from 
the dependence of various quantities on the system size),
rather than suppress them.

We simulate the finite-volume system (cubic box, periodic
boundary conditions) directly at the critical point, and
study the probability distributions and fluctuation matrices
of various observables. Our method has much in common with,
and can be viewed as a generalization of, the methods
used in \cite{wilding,alonso}. The details of our study
can be found in \cite{endpoint}. Our main results are:

1. The location of the endpoint corresponds, after
extrapolation to the continuum limit, to the value of the Higgs
mass in the Standard Model $m_{H,c} = 72 \pm 2$ GeV. 
As experimentally $m_H > 88$ GeV \cite{mHlower}, the phase
transition in the Standard Model does not occur.

2. We have obtained strong evidence that the endpoint of the 
electroweak phase transition belongs to the universality 
class of the 3d Ising model. The $M$-like and $E$-like
observables that we have identified at the critical point
of the model (\ref{latticeaction}), demonstrate the joint
probability distribution that matches very well the joint
distribution of magnetization and energy of the 3d Ising model
(\fig\ref{2dplots}). The matching improves with larger
lattices and larger space of observables. (Such distributions 
can serve as a very sensitive indicator of the universality 
class: the corresponding distributions for O(2) and O(4) spin
models are very different, as shown in the same figure).
The critical indices computed from the finite size 
scaling relations are also consistent with the 3d Ising model 
universality class.

\section*{Acknowledgments}

The work reported here was done in collaboration with K. Kajantie, 
M. Laine, K. Rummukainen and M. Shaposhnikov~\cite{endpoint}. 
This work 
was partly supported by the TMR network {\em Finite Temperature Phase
Transitions in Particle Physics}, EU contract no.\ FMRX-CT97-0122,
and by the Russian Foundation for Basic Research.

\end{document}